\documentstyle[11pt,epsf]{article}
\def\Journal#1#2#3#4{{#1}{\bf #2}, #3 (#4)}

\def\NPB{{\em Nucl. Phys.} B}
\def\PLB{{\em Phys. Lett.} B}

\def\PRD{{\em Phys. Rev.} D}
\def\RMP{\em Rev. Mod. Phys. }
\def\SJNP{\em Sov. J. Nucl. Phys. }

\def\be{\begin{equation}}
\def\ee{\end{equation}}
\def\bea{\begin{eqnarray}}
\def\eea{\end{eqnarray}}

\begin{document}
\thispagestyle{empty}
\setcounter{page}0

{\hbox to\hsize{August, 1996  \hfill TAC-1996-018}}

~\vfill
\begin{center}
{\Large\bf    Nonequilibrium Neutrino Oscillations\\
                            and\\
              Primordial Production of $^4\! He$
}
\vfill
{\large D. P. Kirilova}
\vspace{.5cm}

{\em
Teoretisk Astrofysik Center\\
Juliane Maries Vej 30, DK-2100, Copenhagen, Denmark
\footnote{Permanent address:
{\em Institute of Astronomy, Bulgarian Academy of Sciences,\\
blvd. Tsarigradsko Shosse 72, Sofia, Bulgaria}\\
  }
}\\

\vspace{.5cm}
and \\

\vspace{1cm}

{\large M. V. Chizhov}
\vspace{.5cm}

{\em
Centre for Space Research and Technologies, Faculty of Physics,\\
University of Sofia, 1164 Sofia, Bulgaria\\
E-mail: $mih@phys.uni$-$sofia.bg$}

\end{center}
\vfill

\begin{abstract}
We studied nonequilibrium oscillations between left-handed
electron neutrinos and nonthermalized sterile neutrinos
in medium for resonant and nonresonant cases.
The exact kinetic equations for neutrinos,
written in terms of neutrino density matrix
in {\it momentum space} were analyzed.
The evolution of neutrino density matrix was
numerically calculated. This allowed
to study precisely the evolution of the neutrino number densities,
energy spectrum distortion and the asymmetry between neutrinos and
antineutrinos for each momentum mode.
The influence of nonequilibrium oscillations on the primordial production
of $^4\! He$ was calculated. Cosmological constraints on the oscillation
parameters were obtained.
\end{abstract}

\vfill

\newpage

\section{Nonequilibrium neutrino oscillations}

   We discuss nonequilibrium oscillations between weak interacting
electron neutrinos $\nu_e$ and sterile neutrinos $\nu_s$
for the case when $\nu_s$ do not thermalize till 2 $MeV$
and oscillations become effective after $\nu_e$ decoupling.
Oscillations of  that type, but for the case of $\nu_s$ thermalizing
before or around 2 $MeV$ have been already discussed in
literature [1-8]. We have provided a proper kinetic
analysis of the neutrino evolution in terms of {\it kinetic equations
  for the neutrino density matrix in momentum space}. Such approach,
though much more complicated, reveals some beautiful features of
 noneqilibrium oscillations, that cannot be catched otherwise.

The assumptions of the model are the following:
\begin{itemize}
\item
Singlet neutrinos decouple much earlier than
the active neutrinos do:  $T^F_{\nu_s} \ge T^F_{\nu_e}$,
as far as they do not participate in the
ordinary weak interactions.
\end{itemize}
Therefore, in later epochs their
temperature and number densities are considerably less than
those of the active neutrinos due to the subsequent annihilations
and decays of particles that have additionally heated $\nu_e$
in equilibrium in comparison with the already decoupled $\nu_s$.
\begin{itemize}
\item
We consider oscillations between the sterile $\nu_s$ and the active
$\nu_e$ neutrinos, according to the Majorana\&Dirac ($M\&D$)
mixing scheme [9].
\end{itemize}
 For simplicity in what follows we accept a simple
$M\&D$-mixing scheme with mixing present just in the electron sector
$\nu_i={\cal U}_{il}~\nu_l$, $l=e,s$:
\footnote{The transitions between different
neutrino flavours were proved to have negligible effect on the
neutrino number densities and on nucleosynthesis because of the
very slight deviation
from equilibrium in that case $T_f \sim T_f'$ ($f$ is the flalour
index) [10].}
$$
\begin{array}{ccc}
\nu_1 & = & c\nu_e+s\nu_s\\
\nu_2 & = & -s\nu_e+c\nu_s,
\end{array}
$$
where $\nu_s$ denotes the sterile electron antineutrino, $c=cos(\vartheta)$
$s=sin(\vartheta)$ and $\vartheta$ is the mixing angle in the electron sector,
 the mass eigenstates $\nu_1$ and  $\nu_2$ are Majorana particles with masses
correspondingly $m_1$ and $m_2$.

When transitions between active and sterile neutrinos are allowed,
sterile neutrinos may not decouple from the plasma earlier enough
in comparisson with the active ones, so that their densities may not
differ enough for a noneqilibrium to be considerable,
or they may regain their thermal equilibrium if already
decoupled [11, 1-4]. The reactions of $\nu_e$ with the plasma
are the source of thermalization for $\nu_s$,  because when oscillating
into  active neutrinos they may interact with the medium and thus
thermalize. Therefore,
\begin{itemize}
\item
we assume that neutrino oscillations become
effective after the decoupling
of the active neutrinos, $\Gamma_{osc}\ge H$ for $T\le 2 MeV$.
\end{itemize}
This puts constraint on the neutrino mass difference:
 $\delta m^2 \le 1.3 \times 10^{-7}$ $eV^2$.
\begin{itemize}
\item
Finally we require that sterile neutrinos have not thermalized till
2 $MeV$ when oscillations become effective.
\footnote{For the estimation of this rate we have used the pioneer
work of Manohar [11] and also recent publications
on the problem of thermalization of neutrinos by
Barbieri\&Dolgov [1], Kainulainen [2]  and
Enqvist et al. [4].}
\end{itemize}
 This puts the following limit on the allowed
range of oscillation parameters:
 $sin(2\vartheta) \delta m^2 \le 10^{-7} eV^2$.

 As far as for this model
the rates of expansion of the Universe, neutrino oscillations and
neutrino interactions with the medium may be comparable, we
have used kinetic equations for neutrinos accounting {\it simultaneously}
for the participation of neutrinos into expansion, oscillations and
interactions with the medium.
Moreover, as far as nonequilibrium oscillations are concerned,
the density matrix of neutrinos considerably differs from its
equilibrium form, and therefore one should work in terms of
density matrix of neutrinos in momentum space [10, 12, 13].
\footnote{When neutrinos are in equilibrium their density matrix
has its equilibrium form, namely $\rho_{ij}=\delta_{ij} \exp(\mu/T-E/T)$,
so that one can work with particle densities instead of $\rho$.
In an equilibrium background, the introduction of oscillations
slightly shifts $\rho$ from its diagonal form, due to the extreme
smallness of the neutrino mass in comparison with the characteristic
temperatures and to the fact that equilibrium distribution of massless
particles is not changed by the expansion [10].}
We have analyzed the evolution of nonequilibrium oscillating
neutrinos by numerically integrating the kinetic equations for the
density matrix in {\it momentum} space for the period after the decoupling
of the electron neutrino till the freezing of neutron-proton ratio
($n/p$-ratio), i.e. for $2~ MeV \ge T \ge 0.3~ MeV$.
 We considered both resonant $\delta m^2 = m_2^2 - m_1^2 <0$
 and nonresonant $\delta m^2 >0$ neutrino oscillations.

Three main effects of neutrino nonequilibrium oscillations were
revealed and precisely studied:

(a) As far as oscillations become effective when
the number densities of $\nu_e$
are much greater than those of $\nu_s$, $N_{\nu_e} \gg N_{\nu_s}$,
the oscillations tend to reestablish the statitstical
eqilibrium between different oscillating species.
As a result $N_{\nu_e}$ will decrease in comparison to its standard
eqilibrium value. (This result of our study is in accordance with
other publications concerning depletion of electron neutrinos [3-7]).

(b) For the case of strongly noneqilibrium oscillations
the distortion of the energy distribution of neutrinos may be
considerable and must be accounted for properly.\footnote{This
effect was discussed for a first time in Ref.~[10],
but unfortunately, as far as the case of flavour neutrino
oscillations were considered and the energy distortion for that case was
shown to be negligible, it was not paid the necessary attention it
deserved.} This effect was not discussed in publications concerning
active-sterile neutrino oscillations, and was thought to be negligible.
In Ref.~[12] it was first shown that for the case of
$\nu_e \leftrightarrow \nu_s$ {\it vacuum oscillations} this effect
is considerable and may be even greater than that of an additional
neutrino species. In the present
work we discuss this effect for neutrino {\it oscillations
in a medium}. We have studied the evolution of the energy distortion using
kinetic equations for neutrino density matrix in
{\it momentum space} [10, 12, 13].
 As far as these integro-differential equations are coupled
nonlinear equations,  analytic solution cannot be obtained
without drastic assumptions. Therefore, we have numerically explored the
problem.

The evolution of the distortion is the following:
First the low energy part of the spectrum is distorted, and later on
this distortion concerns neutrinos with higher and higher
energies (Fig.~1). This behavior is
natural, as far as neutrino oscillations affect first low energy neutrinos,
$\Gamma_{osc} \sim \delta m^2/E_{\nu}$.
 The naive account of this effect by shifting the effective temperature and
assuming the neutrino spectrum of equilibrium form gives wrong results
for the case $\delta m^2 < 10^{-7} eV^2$.
For bigger neutrino mass differences
oscillations are fast enough and the naive account looks more acceptable,
provided that $\nu_e$ have not decoupled.

(c) Other interesting effect revealed by our approach is the
generation of asymmetry between $\nu_e$ and their antiparticles.
The possibility of an asymmetry generation
due to CP-violating
flavour oscillations was first proposed in Ref.~[14].
Later estimations
of an asymmetry due to CP-violating MSW resonant oscillations
were provided [15]. Recently the problem of asymmetry was
considered in connection with the exploration of the neutrino
propagation in the CP-odd plasma of the early Universe [1-5]
and this type of asymmetry was obtained to be negligible.
Recently it was shown in Ref.~[16, 7], that
asymmetry can grow to a considerable values for
the case of great mass differences, $\delta m^2 \ge 10^{-5} eV^2$ was
discussed.
Our approach allows precise description of the asymmetry and its
evolution, as far as working with the {\it selfconsistent
kinetic equations
for neutrinos in momentum space} enabled us to calculate the
behavior of asymmetry at each momentum. So, the calculated
result may differ considerably from the rough
estimations made by working with neutrino mean energy and with the
integrated quantities like particle densities and
the energy densities.

For the nonresonant case
we obtained that the asymmetry is negligible.
The asymmetry effect is noticeable only
for the resonant case, as it was expected. Even when the asymmetry
is assumed initially negligibly small (of the order of the baryon one),
i.e. $\sim 10^{-10}$,
it may be considerably amplified at resonant transition due to different
interactions of neutrinos and antineutrinos with the CP-odd medium.
 The value of the asymmetry may increase
by several orders of magnitude, oscillating and sign changing.
This behavior may remain such long after the resonant transition.
(This behavior needs very high accuracy and thus complicates strongly
the numerical analysis.) Even in case when the value of the asymmetry
does not become considerable enough to have some direct noticeable effect,
on primordial nucleosynthesis for example,
the asymmetry term at the resonant transition
 determines the evolution of the neutrino density matrix.
 (As far as the leading order terms, namely of oscillation effects and
 interaction with the medium, compensate each other at
 resonant transition.) It effectively suppresses the transitions of active
neutrinos (antineutrinos)
thus weakening the effect of neutrino depletion at resonance.
 Thus the evolution of neutrino may drastically differ from
 the case without  asymmetry. For some model parameters this effect
consists 20\% of the previously discussed. We will discuss asymmetry
evolution in more detail elsewhere. Here finally we
want only to note, that linear approximation for the asymmetry in the
resonant case may give very misleading results.

As an illustration of the strength and importance of these effects we
discuss their influence on the
primordial production of $^4\! He$.
Effect of oscillations on nucleosynthesis has been discussed in
numerous publications [1-8, 12, 14]. However, a detail kinetic
calculations of helium abundance for the case of nonequilibrium
oscillations in medium has not been done.  We describe it in the
last section. Here we briefly summerize the results:

The depletion of the electron neutrino number densities due to
oscillations to sterile ones leads to an effective decrease in the
weak processes rates, and thus to an increase of the freezing
temperature of the $n/p$-ratio and corresponding overproduction of the
primordially produced $^4\! He$. The effect of the distortion of the
energy distribution of neutrinos has two aspects. An average
decrease of the energy of active neutrinos should lead to a
decrease of the weak reactions rate  , $\Gamma_w \sim E_\nu^2$ and
subsequently to an increase in the freezing temperature and the
produced helium. On the other hand, there exists an energy
thershold for the reaction $\tilde{\nu_e}+p \to n+e^+$. And in
case when, due to oscillations, the energy of the relatively
greater part of neutrinos becomes smaller than that threshold
the $n/p$- freezing ratio decreases [17].
However, the numerical analysis showed that the latter effect is
less noticeable compared with the previously described ones.
The asymmetry calculations showed a slight predominance of neutrinos
over antineutrinos, leading to decrease of helium. The greater effect
of the asymmetry is, however, the relative increase of both neutrino
and antineutrino particle densities  which leads to a noticeable
underproduction of helium (up to $10\%$ relative decrease)
Thus, the total result of nonequilibrium neutrino oscillations is
overproduction of helium in comparison to the standard value.
We have used the $4\%$ relative increase in the primordially produced
helium to obtain the exclusion region for the oscillation parameters.

 \section{The kinetics of noneqilibrium neutrino oscillations}

The kinetic equations for the density matrix of the noneqilibrium
oscillating neutrinos in the primeval plasma of the Universe
in the epoch previous to nucleosynthesis, i.e. consisting of
photons, neutrinos, electrons, and small quantities of nucleons,
have the form:
\be
{\partial \rho(t) \over \partial t} =
H p~ {\partial \rho(t) \over \partial p}
+ i \left[ {\cal H}_o, \rho(t) \right]
+i \left[ {\cal H}_{int}, \rho(t) \right]
+ {\rm O}\left({\cal H}^2_{int} \right),
\label{kin}
\ee
where $p$ is the momentum of electron neutrino and
$\rho$ is the density matrix of the massive Majorana
neutrinos in momentum space.

The first term in the equation describes the effect of expansion,
the second is responsible for oscillations, the
third accounts for forward neutrino scattering off the
medium. ${\cal H}_o$ is the free neutrino hamiltonian:
$$
{\cal H}_o = \left( \begin{array}{cc}
\sqrt{p^2+m_1^2} & 0 \\ 0 & \sqrt{p^2+m_2^2}
\end{array} \right),
$$
while ${\cal H}_{int} = \alpha~V$ is the interaction hamiltonian,
where $\alpha_{ij}=U^*_{ie} U_{je}$,
$V=G_F \left(\pm L - Q/M_W^2 \right)$,
and in the interaction basis plays
the role of induced squared mass for electron neutrinos:
$$
{\cal H}_{int}^{LR} = \left( \begin{array}{cc}
V & 0 \\ 0 & 0 \end{array} \right).
$$
The first `local' term in $V$ accounts
for charged- and neutral-current
tree-level interactions with medium protons, neutrons,
electrons and positrons, neutrinos and antineutrinos.
It is proportional to
the fermion asymmetry of the plasma $L=\sum_f L_f$, which is assumed
of the order of the baryon one
$$
L_f \sim {N_f-N_{\bar{f}} \over N_\gamma}~T^3 \sim
{N_B-N_{\bar{B}} \over N_\gamma}~T^3 = \beta T^3.
$$
 The second `nonlocal' term arises as an $W/Z$ propagator effect,
$Q \sim E_\nu~T^4$.
For the early Universe conditions both terms must be accounted for
because although the second term is of the second power of $G_F$ ,
the first term is proportional besides to the first power
of $G_F$, to the very small value of the fermion
asymmetry (of the order of the baryon asymmetry, i.e. $10^{-10}$
in case $B-L$ conservation is accepted) [18].
Moreover, the two terms have different
temperature dependence and an interesting interplay between them
during the cooling of the Universe is observed.

The last term in the Eq.~(\ref{kin}) describes the weak interactions
of neutrinos with the medium. For example, for the weak reactions
of neutrinos with electrons and positrons $e^+ e^- \leftrightarrow
\nu_i \tilde{\nu}_j$, $e^\pm \nu_j \to e'^\pm \nu'_i$
it has the form
$$
\begin{array}{cl}
 & \int {\rm d}\Omega(\tilde{\nu},e^+,e^-)\left[
n_{e^-} n_{e^+} {\cal A} {\cal A}^\dagger - \frac{1}{2} \left\{
\rho,~ {\cal A}^\dagger \bar{\rho} {\cal A} \right\}_+ \right] \\
+ & \int {\rm d}\Omega(e^-,\nu',e'^-)\left[
n'_{e^-} {\cal B} \rho' {\cal B}^\dagger - \frac{1}{2} \left\{
{\cal B}^\dagger {\cal B}, ~\rho \right\}_+ n_{e^-} \right] \\
+ & \int {\rm d}\Omega(e^+,\nu',e'^+)\left[
n'_{e^+} {\cal C} \rho' {\cal C}^\dagger - \frac{1}{2} \left\{
{\cal C}^\dagger {\cal C}, ~\rho \right\}_+ n_{e^+} \right],
\end{array}
$$
where $n$ stands for the number density of
the interacting particles,
${\cal A}$ is the amplitude of the process
$e^+ e^- \to \nu_i \tilde{\nu}_j$,
${\cal B}$ is the amplitude of the process
$e^- \nu_j \to e'^- \nu'_i$ and
${\cal C}$ is the amplitude of the process
$e^+ \nu_j \to e'^+ \nu'_i$.
They are expressed through the known amplitudes
${\cal A}_e(e^+ e^- \to \nu_e \tilde{\nu}_e)$,
${\cal B}_e(e^- \nu_e \to e^- \nu_e)$ and
${\cal C}_e(e^+ \nu_e \to e^+ \nu_e)$:
$$
{\cal A} = \alpha~{\cal A}_e,~~~~~
{\cal B} = \alpha~{\cal B}_e,~~~~~
{\cal C} = \alpha~{\cal C}_e.
$$
We have analysed the evolution of the neutrino density matrix
assumed that oscillations become noticeable
after electron neutrinos decoupling. So, the neutrino kinetics
down to 2 $MeV$ does not differ from the standard case, i.e.
electron neutrinos maintain their equilibrium distribution,
while sterile neutrinos are absent.
Then the last term in the kinetic equation can
be neglected.
The equation results into a set of coupled nonlinear
integro-differential equations for the components
of the density matrix.
Analytical solution is not possible.\footnote{For the case
of vacuum neutrino oscillations
this equation was analytically solved and the
evolution of density matrix was given explicitly in Ref.~[12].}

We have numerically calculated the evolution of the
neutrino density matrix for temperature interval
$[0.3,2.0]~ MeV$, i.e. after neutrino decoupling and till
the $n/p$ freezing. The oscillation parameters range
studied is $\delta m^2 \in \pm [10^{-10}, 10^{-7}]~eV^2$
 and $\vartheta \in[0,\pi/4]$.
The baryon asymmetry $\beta$ was taken to be $3\times 10^{-10}$.

As was already stated our numerical analysis showed that
the nonequilibrium oscillations
can considerably deplete the number densities of
electron neutrinos (antineutrinos) and distort their
energy spectrum. Neutrino-antineutrino asymmetry
may grow at the resonant transition and may also effect
considerably the evolution
of neutrino ensembles.

\section{Nucleosynthesis with nonequilibrium oscillating neutrinos}

The kinetic equation describing the evolution of the neutron number
density in momentum space $n_n$ for the case of oscillating neutrinos
$\nu_e \leftrightarrow \nu_s$ reads:
\be
\begin{array}{cl}
{\partial n_n \over \partial t} = H p_n~ {\partial n_n \over \partial p_n}
& + \int {\rm d}\Omega(e^-,p,\nu) |{\cal A}(e^- p\to\nu n)|^2
(n_{e^-} n_p - n_n \rho_{LL}) \\
& - \int {\rm d}\Omega(e^+,p,\tilde{\nu}) |{\cal A}(e^+n\to p\tilde{\nu})|^2
(n_{e^+} n_n - n_p \bar{\rho}_{LL}).
\end{array}
\ee
The first term describes the expansion and the second one -- the
processes $e^- + p \leftrightarrow n + \nu_e$ and
$p + \tilde{\nu}_e \leftrightarrow e^+ + n$,
directly influencing the nucleon density.
It differs from the standard scenario one only by the substitution
of $\rho_{LL}$ and $\bar{\rho}_{LL}$ instead of
$n_{\nu}^{eq} \sim \exp(-E_\nu/T)$.
The neutrino and antineutrino density matrices may differ
$\bar{\rho}_{LL} \ne \rho_{LL}$, contrary
to the standard model, as a result of the different reactions
with the CP-odd plasma of the prenucleosynthesis epoch.
Number densities per unit volume are expressed as
$N = (2\pi)^{-3}\int{\rm d}^3 p~n(p)$.

We have numerically integrated the equation
for the temperature range of interest $T \in [0.3,2.0]~ MeV$ for the
range of oscillation parameters of our model. The results of the
numerical integration are illustrated in Fig.~2,
in comparison with the vacuum case and the standard nucleosynthesis
without oscillations.
It is obvious that the kinetic effects of the nonequilibrium
oscillations on nucleosynthesis are considerable.
The next figure presents the dependence of the primordially
produced helium on mixing angle (Fig.~3).

From numerical integration for different oscillation parameters
we have obtained constant $^4\! He$ contours.
The Fig.~4 gives the cosmologically excluded regions of
oscillation parameters. It corresponds to primordial abundance
of helium $Y_p=0.245$, which gives $4\%$  overproduction of helium
in comparison with the observational value.

For the cases when the energy
distortion and asymmetry are considerable we have obtained an
order of magnitude stronger constraints than the cited in
literature [4-7]. Therefore, in conclusion we would like to
stress once again,
that in case of nonequilibrium neutrino oscillations working
with the exact kinetic equations for the density matrix of
neutrinos in momentum space is necessary.

\section*{Acknowledgments}
D.K. is grateful to A.D.Dolgov for introducing her into the realm of
nonequilibrium neutrino oscillations and for numerous fruitful discussions.
We thank ICTP Trieste, where part of the numerical analysis was done.
This paper was supported in part by the Danish National Science
Research Council through grant 11-9640-1 and in part by Denmarks
Grundforskningsfond through its support of the Theoretical Astrophysical
Center.

\vspace{1cm}

\pagebreak[1]

\newpage

\begin{center}{\Large Figure Captions}
\end{center}
\ \\

{\bf Figure 1}: The figure shows the energy distortion of active
neutrinos $x^2 \rho_{LL}(x)$, where $x=E_\nu/T$, for the case of
nonequilibrium neutrino oscillations,
$\delta m^2 = -10^{-8}$, $\vartheta=\pi/8$
at different temperatures:
$T=1~MeV$ (a), $T=0.7~MeV$ (b), and $T=0.5~MeV$ (c).

\ \\

{\bf Figure 2}: The curves represent the evolution of the neutron
number density relative to nucleons $X_n(t)=N_n(t)/(N_p+N_n)$ for
the nucleosynthesis model with vacuum nonequilibrium oscillations
and for the case of nonequilibrium oscillations in medium,
$\delta m^2 = -10^{-8}$, $\vartheta=\pi/8$. For comparison the curve
corresponding to the standard nucleosynthesis model is shown.

\ \\

{\bf Figure 3}: The figure illustrates the dependence of the neutron
number density relative to nucleons $X_n=N_n/(N_p+N_n)$ on the
mixing angle for $\delta m^2 = -10^{-8}$.

\ \\

{\bf Figure 4}: Exclusion regions for oscillation parameters are shown
for the case of resonant $\delta m^2 < 0$ and nonresonant $\delta m^2 > 0$
neutrino oscillations. The curves correspond to helium abundance
$Y_p=0.245$.

\pagestyle{empty}

\epsfbox[100 170 700 770]{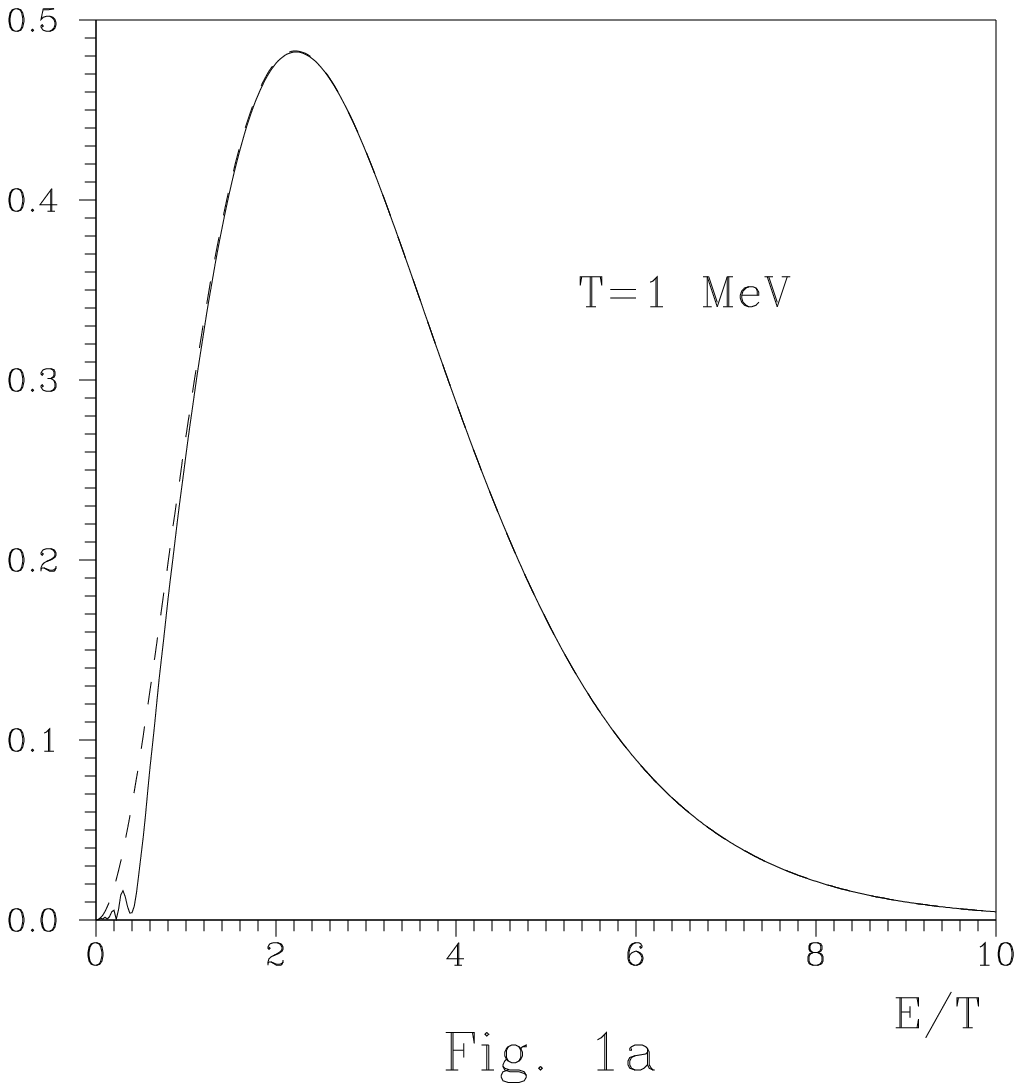}

\newpage

\epsfbox[100 170 700 770]{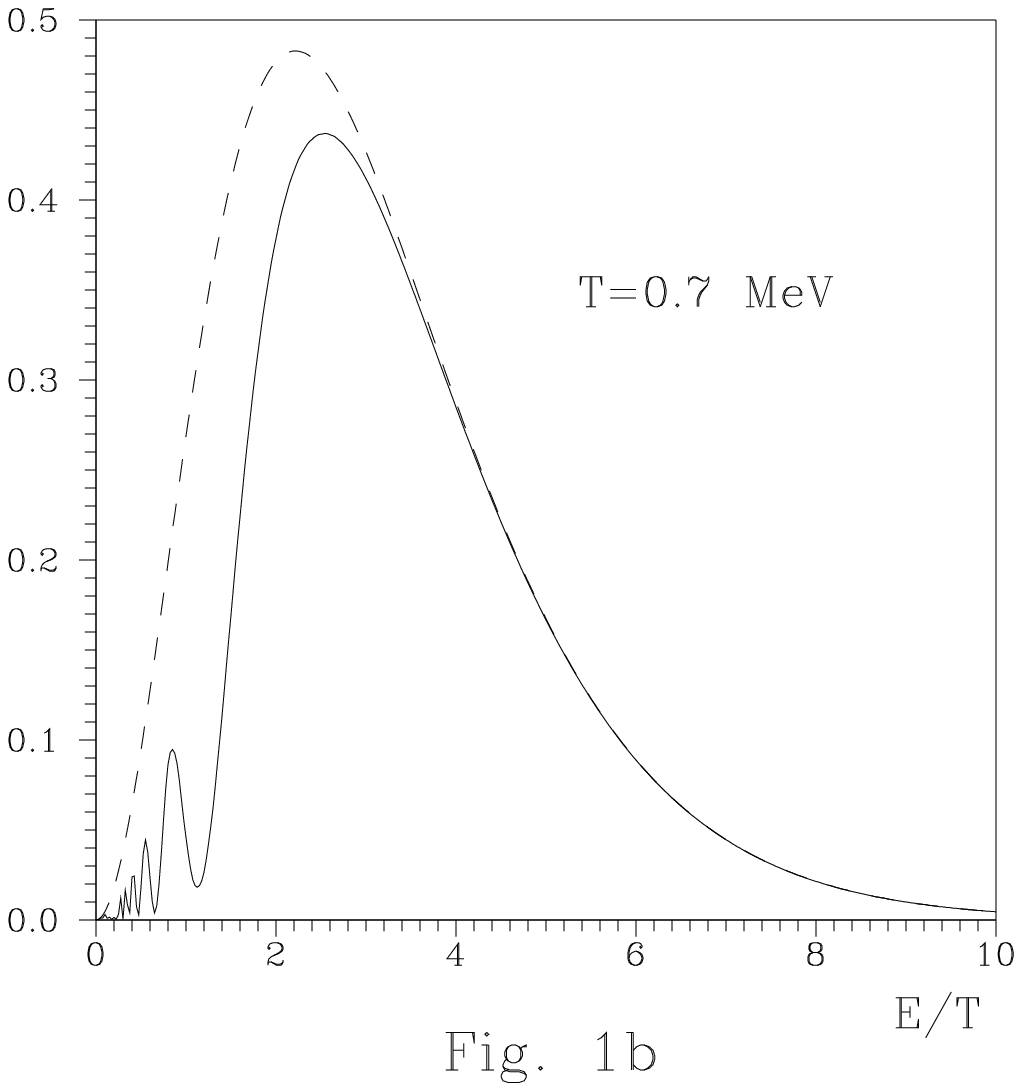}

\newpage

\epsfbox[100 170 700 770]{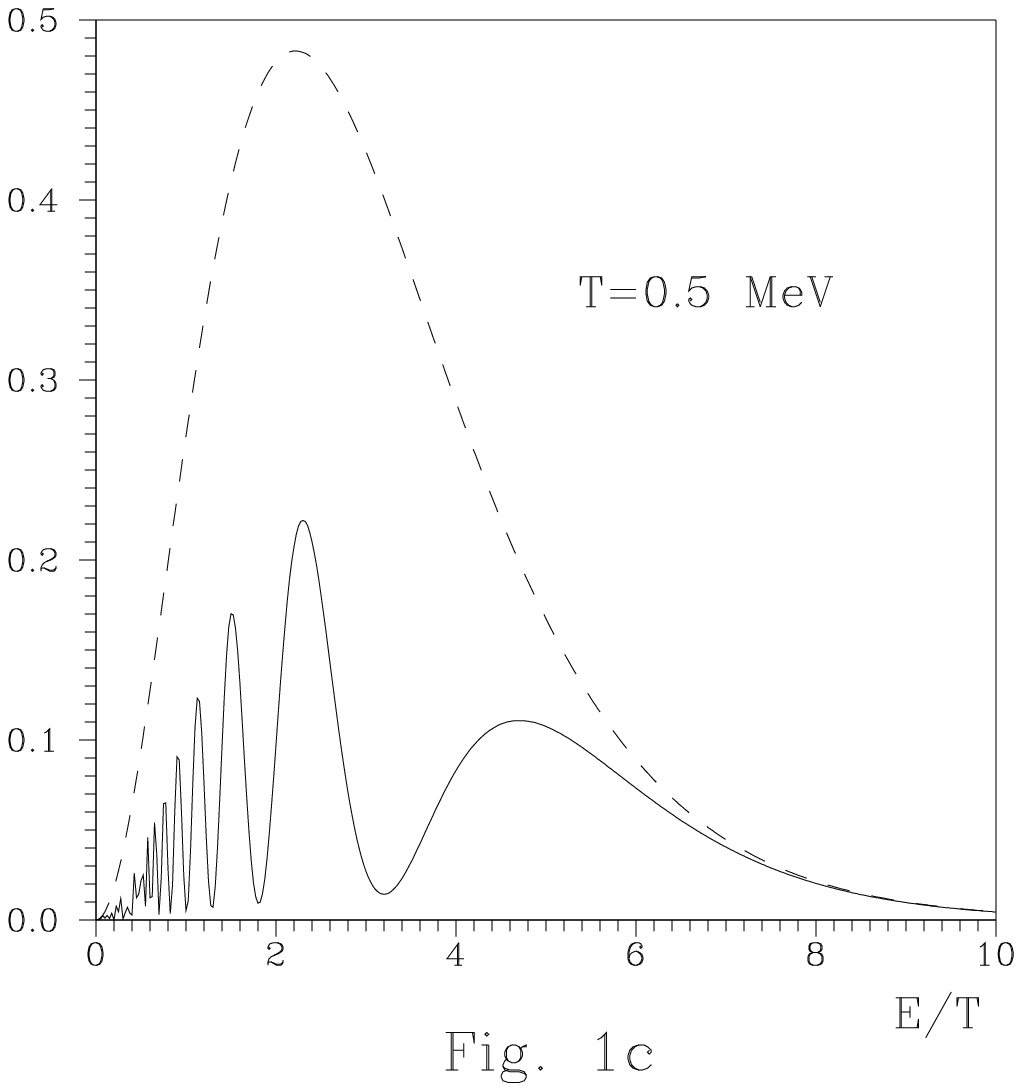}

\newpage

\epsfbox[100 170 700 770]{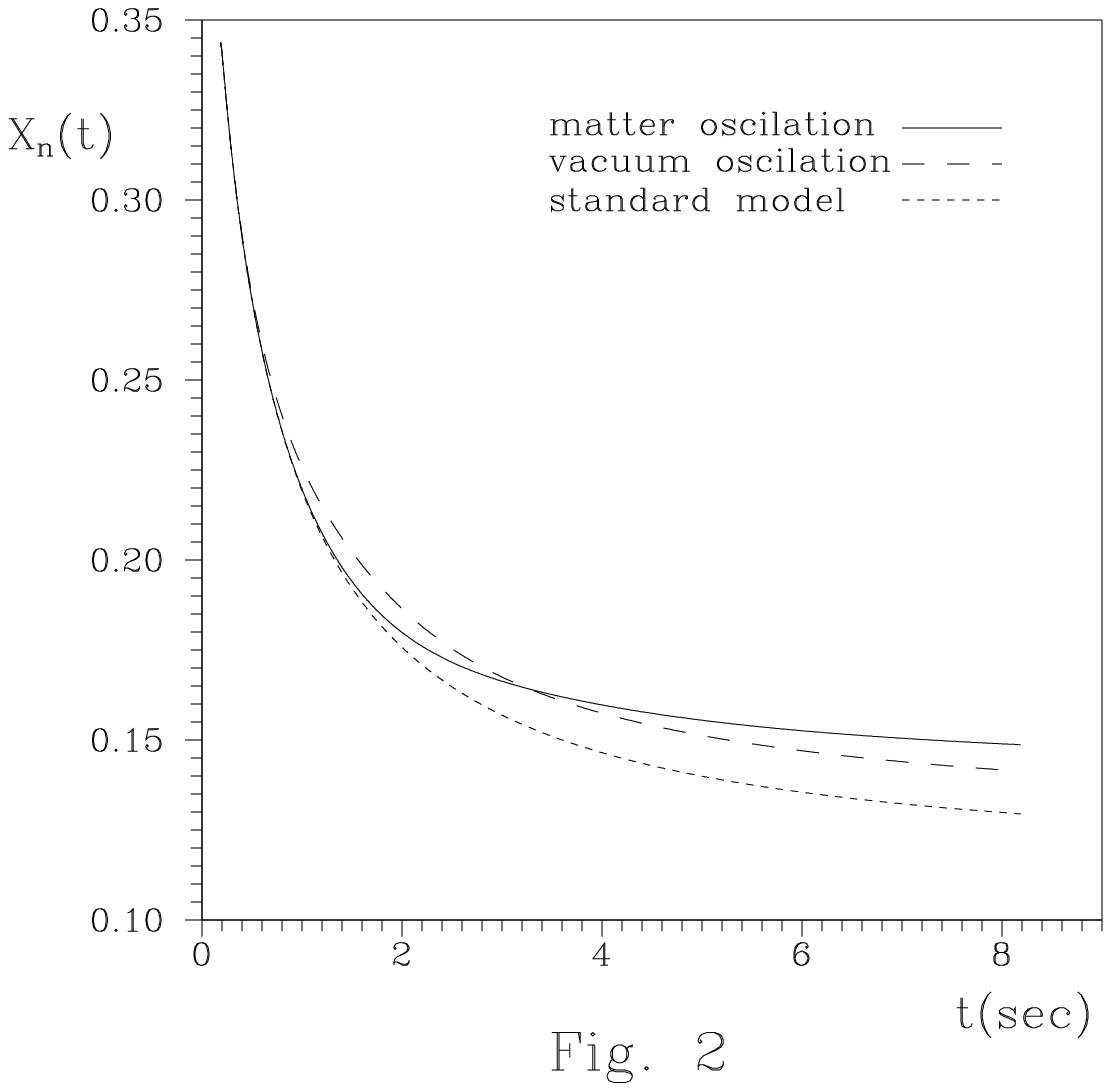}

\newpage

\epsfbox[100 170 700 770]{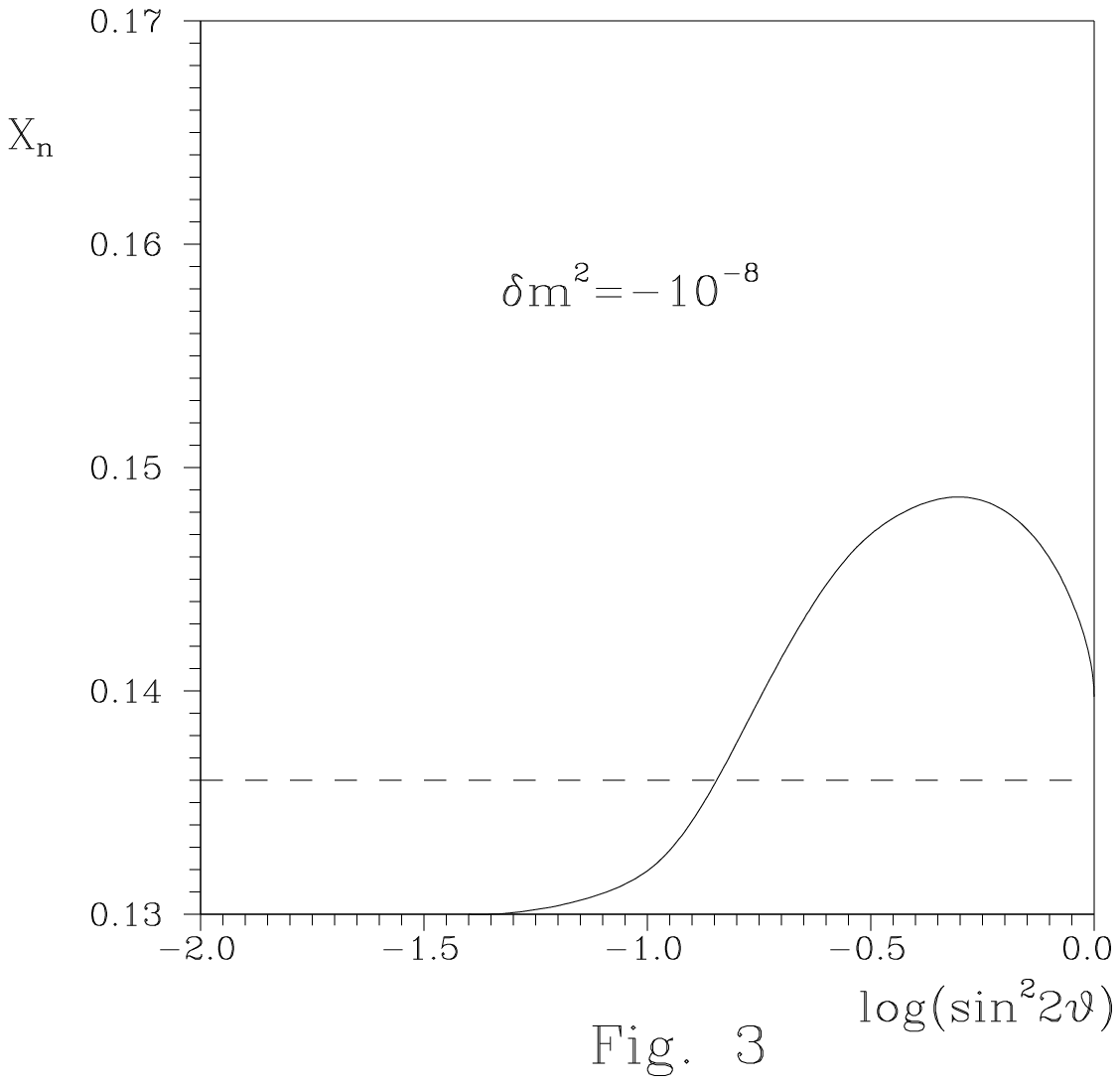}

\newpage

\epsfbox[100 170 700 770]{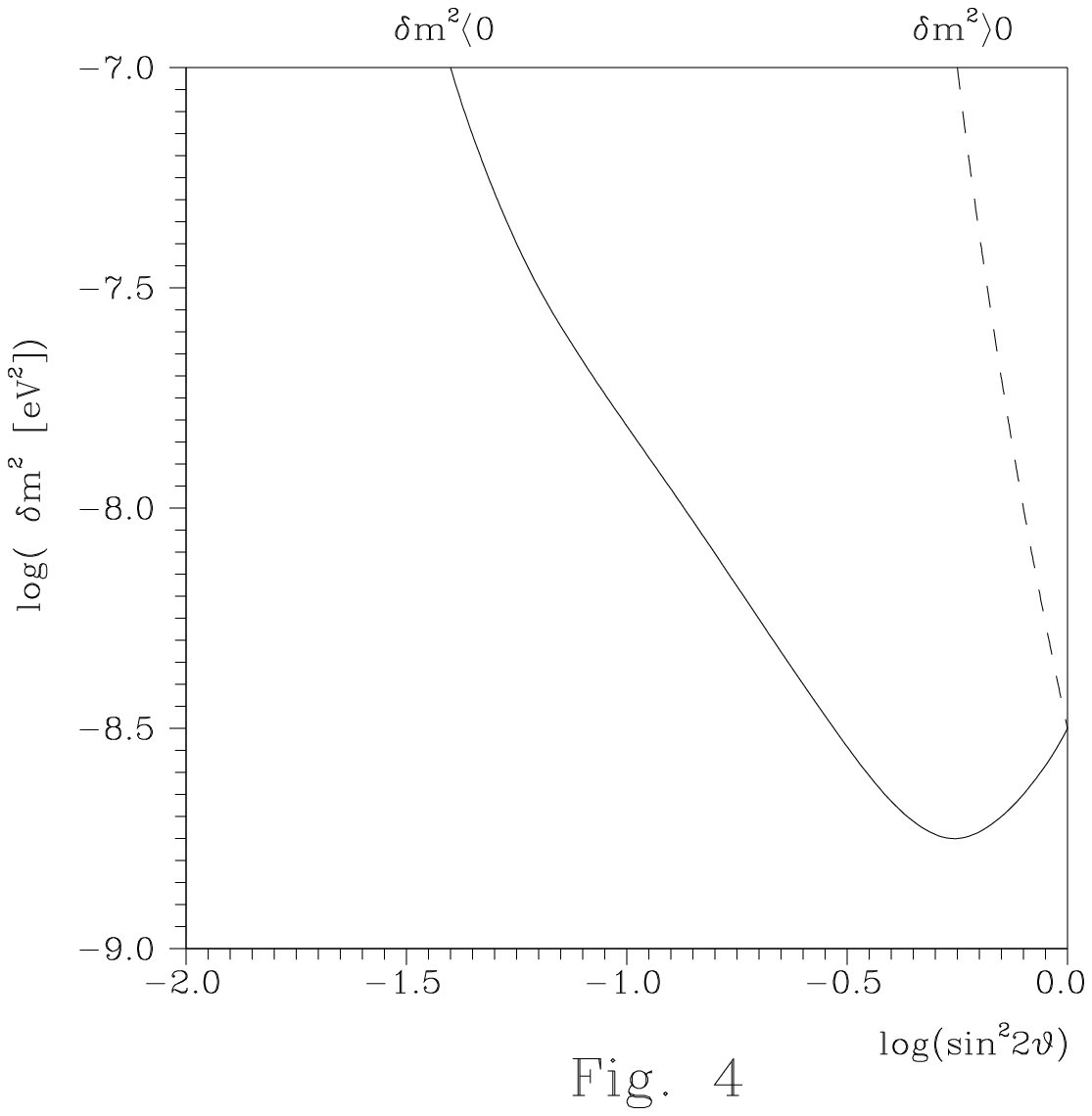}

\end{document}